\documentclass[12pt]{iopart}
\usepackage{graphicx}

\begin{document}

\title[Testing the EEP with two Earth-orbiting clocks]{Testing the Einstein Equivalence Principle with two Earth-orbiting clocks}

\author{Dmitry Litvinov\,$^{1,2}$ and Sergey Pilipenko\,$^1$}

\address{$^1$ Astro Space Center, Lebedev Physical Institute, Profsoyuznaya 84/32, 117997 Moscow, Russia}

\address{$^2$ Bauman Moscow State Technical University, 2-ya Baumanskaya 5, 105005 Moscow, Russia}

\ead{litvirq@yandex.ru}
\vspace{10pt}
\begin{indented}
\item[]8 January 2021
\end{indented}

\begin{abstract}
We consider the problem of testing the Einstein Equivalence Principle (EEP) by measuring the gravitational redshift with two Earth-orbiting stable atomic clocks. For a reasonably restricted class of orbits we find an optimal experiment configuration that provides for the maximum accuracy of measuring the relevant EEP violation parameter. The perigee height of such orbits is $\sim$~1,000~km and the period is 3--5~hr, depending on the clock type. For the two of the current best space-qualified clocks, the VCH-1010 hydrogen maser and the PHARAO cesium fountain clock, the achievable experiment accuracy is, respectively, $1\times10^{-7}$ and $5\times10^{-8}$ after 3 years of data accumulation. This is more than 2 orders of magnitude better than achieved in Gravity Probe A and GREAT missions as well as expected for the RadioAstron gravitational redshift experiment. Using an anticipated future space-qualified clock with a performance of the current laboratory optical clocks, an accuracy of $3\times10^{-10}$ is reachable. 
\end{abstract}

%
\vspace{2pc}
\submitto{\CQG}
%
%
%

\section{Introduction}

The recent progress in the optical clock technology \cite{mcgrew-2018-nat, bothwell-2019-mlg} promises significant improvement of the currently achieved accuracy of gravitational redshift tests of the Einstein Equivalence Principle (EEP) \cite{will-2014-lrr}. The accuracy and stability of the current laboratory optical clocks reached $\sim 10^{-18}$ that allowed ground-based gravitational redshift measurements to reach an accuracy of $\sim 10^{-4}$ \cite{takamoto-2020-natpho}, which is comparable to that of the current best space experiments. All the space experiments performed so far, however, has been realized in suboptimal orbital configurations \cite{vessot-levine-1980-prl, delva-2018-prl, herrmann-2018-prl, litvinov-2018-pla}. With space-qualified versions of the optical clocks expected to appear in the current decade \cite{origlia-2018-pra}, it is of much interest to determine the maximum achievable accuracy of such space-based tests. 

In the simplest case of a weak static gravitational field and a single EEP violation parameter, the expression for the gravitational frequency shift of a signal sent by one stationary clock and received by another is:
\begin{equation}
   \frac{\Delta f_\mathrm{grav}}{f} = (1+\varepsilon) \frac{\Delta U}{c^{2}}.
\label{eq:main-freq-violated}
\end{equation}
where $\Delta f_\mathrm{grav}/f$ is the fractional frequency shift, $\Delta U$ is the gravitational potential difference between the clocks, $c$ is the speed of light, and $\varepsilon$ is the EEP violation parameter \cite{vessot-levine-1979-grg}. No significant deviation of $\varepsilon$ from zero has been detected to date, while the measurement accuracy reached $\sim10^{-5}$ \cite{takamoto-2020-natpho, vessot-levine-1980-prl, delva-2018-prl,herrmann-2018-prl}.

In the space gravitational redshift experiments performed so far, one of the clocks was situated on board a satellite and the other installed at a ground station (Gravity Probe A \cite{vessot-levine-1980-prl}, GREAT \cite{delva-2018-prl,herrmann-2018-prl}, RadioAstron \cite{litvinov-2018-pla}). The upcoming ACES experiment is based on this approach as well \cite{aces-2011-acau}. The principal drawback of this scheme is that the stability and accuracy of the clock signals are degraded during their propagation through the noisy atmosphere. Additional noise is added by mechanical vibrations of the ground antenna. Several approaches based on the use of multiple communications links with the satellite have been developed to reduce these effects but complete elimination of those noises appears to be impossible \cite{smarr-vessot-1983-grg,tinto-1996-prd}. An obvious solution to this problem is to use two spacecraft instead of one, each equipped with a highly stable clock. In this paper we consider the problem of determining the maximum possible accuracy of such tests in the particular case when both spacecraft are Earth satellites.

The paper is organized as follows. In Section~\ref{sec:experiment-model} we present the mathematical equations we use to model gravitational redshift experiments with two satellites, as well as some simplifying assumptions. In Section~\ref{sec:accuracy-estimation-approach} we provide details on how we estimate the experiment accuracy and present the covariance matrices for the colored noise of the clocks we consider. In Section~\ref{sec:results} we present the results of our simulations of the accuracy of gravitational redshift experiments with two satellites. For the sake of comparison with space-to-ground tests, in Section~\ref{sec:space-ground-experiment} we estimate the accuracy of the gravitational redshift experiment with RadioAstron using the same model. We conclude with the discussion of the results in Section~\ref{sec:conclusions}.

\section{The experiment model}
\label{sec:experiment-model}

Equation (\ref{eq:main-freq-violated}) presents the simplest way of how an EEP violation can manifest itself in gravitational redshift experiments. However, it is necessary to take into account the possibility that $\varepsilon$ depends on the gravitational field source type, e.g. the ratio of protons to neutrons in it, and the type of the quantum transition used in the clocks (nuclear, optical, etc.) \cite{altschul-2015-asr}. Moreover, it was shown recently that for experiments in the near-Earth space, despite the absence of the noon-midnight redshift, the gravitational potentials that enter into the expression for the gravitational redshift with the EEP violating coefficients must include the contributions of the Sun, Moon, and other massive bodies of the Solar System \cite{wolf-blanchet-2016-cqg}. Therefore, for experiments in the near-Earth space the most general equation for the gravitational frequency shift is:
\begin{equation}
\fl
\frac{\Delta f_\mathrm{grav}}{f} 
= \frac{U^{(1)}_\mathrm{E} - U^{(2)}_\mathrm{E}}{c^{2}} 
+ \frac{\varepsilon^{(1)}_\mathrm{E} U^{(1)}_\mathrm{E} - \varepsilon^{(2)}_\mathrm{E} U^{(2)}_\mathrm{E}}{c^{2}}
+ \frac{\varepsilon^{(1)}_\mathrm{S} U^{(1)}_\mathrm{S} - \varepsilon^{(2)}_\mathrm{S} U^{(2)}_\mathrm{S}}{c^{2}}
+ \frac{\varepsilon^{(1)}_\mathrm{M} U^{(1)}_\mathrm{M} - \varepsilon^{(2)}_\mathrm{M} U^{(2)}_\mathrm{M}}{c^{2}}
 + \ldots,
\label{eq:redshift-multiparam-general}
\end{equation}
where the upper subscript labels the clocks and the lower, respectively, the gravitational field sources: $\mathrm{E}$ for the Earth, $\mathrm{S}$ for the Sun, $\mathrm{M}$ for the Moon, and the ellipsis denotes similar terms for other massive bodies of the Solar System.

Let us consider the simplified case when all the EEP violation parameters in (\ref{eq:redshift-multiparam-general}) are equal and the two clocks are identical. Further, we allow for a (small) frequency offset between the two clocks but assume it constant, i.e. the frequency drift is negligible. This assumption is usually valid for cesium and optical clocks, provided the environmental conditions are constant. For hydrogen maser and rubidium clocks this assumption is usually not true and in such case the model should be amended with additional terms that depend linearly, and possibly quadratically, on time. This is straightforward to implement but appears unnecessary for the simplified model we consider. With the above assumptions, for the gravitational frequency shift of a signal sent by one satellite and received by the other we have:
\begin{equation}
\fl \frac{\Delta f_\mathrm{grav}}{f} 
= \frac{U^{(1)}_\mathrm{E} - U^{(2)}_\mathrm{E}}{c^{2}} 
+ \varepsilon \bigg(
\frac{ U^{(1)}_\mathrm{E} - U^{(2)}_\mathrm{E}}{c^{2}}
+ \frac{U^{(1)}_\mathrm{S} - U^{(2)}_\mathrm{S}}{c^{2}}
+ \frac{U^{(1)}_\mathrm{M} - U^{(2)}_\mathrm{M}}{c^{2}}
+ \ldots \bigg)
+ \Delta f_0
+ n(t),
\label{eq:redshift-model}
\end{equation}
where $\varepsilon$ is the EEP violation parameter to be determined,
$\Delta f_0$ is an unknown frequency offset between the two clocks (nuisance parameter),
$n(t)$ is the random process that characterizes the relative fractional frequency fluctuations of the two clocks, and
$t$ is the coordinate time. Note that (\ref{eq:redshift-model}) gives only the part of the signal frequency shift that is due to gravitation. In an actual experiment it is necessary to take into account all the other contributions as well, e.g. those due to the satellite motion.

\renewcommand{\arraystretch}{1.5}
\begin{table}[t]
\footnotesize
\centering
\caption{The power spectral density (PSD) of the fractional frequency fluctuations, $\Delta f/f$, of the clocks used in the analysis. The  numerical values are chosen such that the corresponding Allan deviations meet the clock specifications.}
\begin{indented}
\item[]\begin{tabular}{cc}
\br
Clock & PSD \\ 
\mr
VCH-1010 & 
$1.5\times10^{-26} f^{0}
+ 7.0\times10^{-31} f^{-1} +3.5\times10^{-35} f^{-2}$\\
PHARAO &
$5.0\times10^{-27} f^{0}
+ 7.5\times10^{-33} f^{-1}$\\
JILA SrI &
$2.0\times10^{-31} f^{0}
+ 2.1\times10^{-36} f^{-1}$\\
\br
\end{tabular}
\end{indented}
\label{table:clock-psds}
\end{table}

\renewcommand{\arraystretch}{1.5}
\begin{table}[t]
\footnotesize
\centering
\caption{The fixed parameters of the satellite orbits for the arbitrarily selected epoch of 01/01/2030 00:00:00 UTC.}
\begin{indented}
\item[]\begin{tabular}{lrr}
\br
Orbital parameter & Satellite 1 & Satellite 2\\
\mr
Inclination & 0~deg & 0~deg\\
Perigee & 7,500~km & 7,500~km\\
RA of the asc. node & 0~deg & 0~deg\\
Argument of perigee & 0~deg & 180~deg\\
Mean anomaly at epoch & 0~deg & 180~deg\\
\br
\end{tabular}
\end{indented}
\label{table:orbital-parameters}
\end{table}

\begin{figure}
    \centering
    \includegraphics[width=0.7\textwidth]{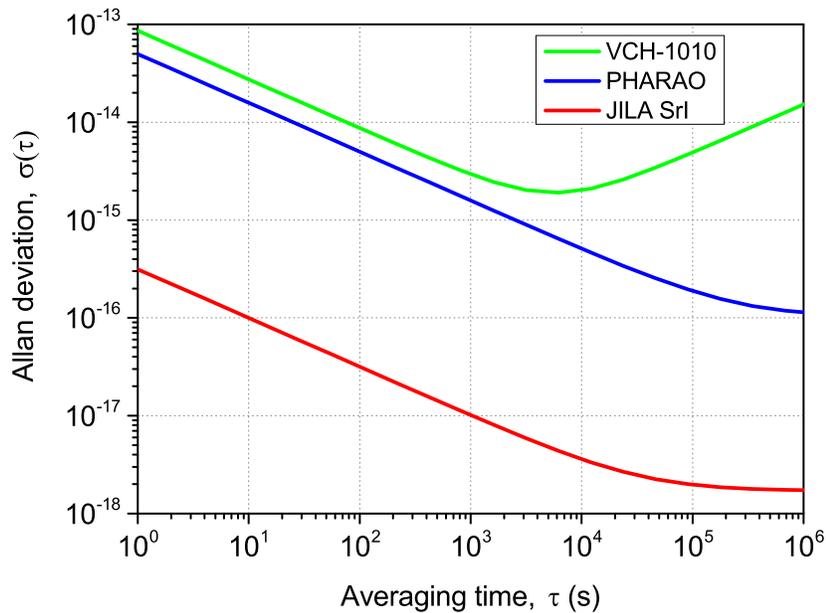}
    \caption{Plots of the Allan deviation (ADEV) of the three clocks: the VCH-1010 hydrogen maser, PHARAO cesium fountain (both space-qualified), and JILA SrI optical laboratory clock. These ADEVs are computed from the PSDs listed in Table~\ref{table:clock-psds} and closely approximate the clock specifications. The increase of the ADEV of VCH-1010 for large averaging times is due to the Brownian noise.}
        \label{fig:adevs}
\end{figure}

Equation (\ref{eq:redshift-model}) constitutes our model of the experiment.
Now, we need to specify the noise process, $n(t)$, and the satellite orbits. We consider the following three types of atomic clocks: the VCH-1010 hydrogen maser used by the RadioAstron spacecraft \cite{litvinov-2018-pla}, the PHARAO cesium fountain clock of the ACES experiment to be performed at the International Space Station \cite{aces-2011-acau}, and the JILA SrI laboratory strontium clock \cite{bothwell-2019-mlg}. Table~\ref{table:clock-psds} lists the expressions for the power spectral density (PSD) of the fractional frequency fluctuations of these clocks that we reconstructed from the available clock specifications. The latter usually specify the types of noise present in the clock signal and the Allan deviation (ADEV) of the clock's frequency fluctuations. In general, it is not possible to reconstruct the PSD from the corresponding ADEV. However, for a mixture of white ($f^{0}$), flicker ($f^{-1}$), and Brownian ($f^{-2}$) frequency noise this is possible unambiguously. The Allan deviation plots that correspond to the PSDs given in Table~\ref{table:clock-psds} are presented in Fig.~\ref{fig:adevs}.

We assume the satellite orbits to be Keplerian. This excludes low Earth orbits from our analysis, since they are influenced by the atmospheric drag on the satellite. This restriction is reasonable since the very idea of the two-satellite experiment is to avoid transmitting signals through the atmosphere. Further, we do not consider the case of high Earth orbits. A large class of such orbits require taking into account the gravitational pull of the Moon and therefore are not Keplerian. Moreover, as will be clear from what follows, for orbits with periods larger than $\sim$~day the experiment accuracy decreases. We also ignore the solar radiation pressure and other non-gravitational forces that influence the satellite motion. Therefore, each orbit we consider can be characterized by a set of 6 Keplerian elements and the experiment accuracy, $\sigma_\varepsilon$, depends on 12 parameters.

We defer the analysis of the full problem in the 12-dimensional parameter space to a future publication and here consider only its subspace, such that the orbits of the two satellites have equal periods and perigees (and thus also apogees). In this case, in order to maximize the amplitude of the gravitational redshift modulation, we need to consider only those pairs of orbits for which the moments of perigee passages by one satellite coincide with the apogee passages by the other. Further, since we do not consider high Earth orbits, the terms due to the Sun, Moon and other planets in (\ref{eq:redshift-multiparam-general}) are several orders of magnitude smaller than those due to the Earth. The dependence of the experiment accuracy on the inclination, right ascension of the ascending node, and the argument of perigee will thus be small and we neglect it. The mean anomaly for one of the satellites can be chosen arbitrarily, we set it to 0$^\circ$. For the other satellite, in order to realize the desired synchronicity of perigee and apogee passages, the mean anomaly has to be set to 180$^\circ$. With the assumptions made, the epoch for the elements can be chosen arbitrarily and we set it to 01/01/2030 00:00:00~UTC.

\begin{figure}
    \centering
    \includegraphics[width=0.7\textwidth]{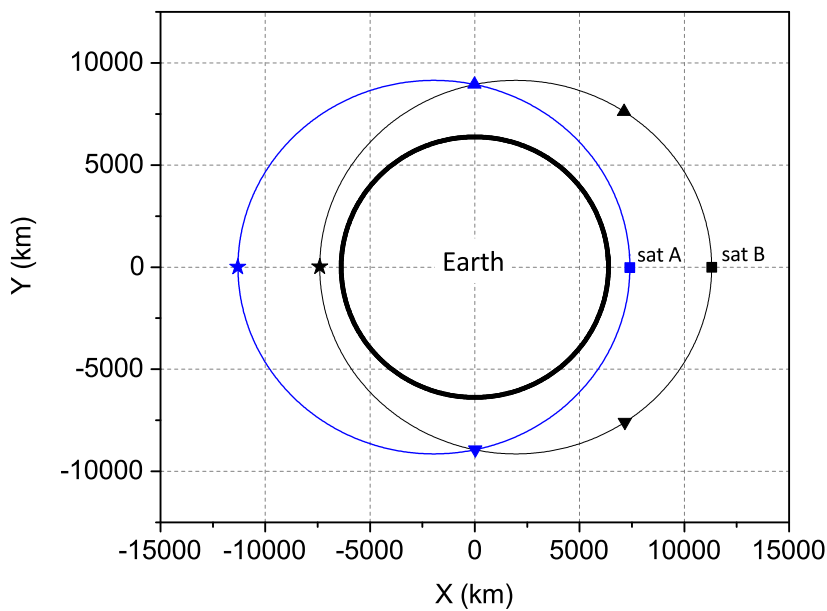}
    \caption{The configuration of the experiment to test the Einstein Equivalence Principle with two satellites that corresponds to the orbital parameters specified in Table~\ref{table:orbital-parameters} and a period of 2.5~hr. The identical symbols mark the positions of the two spacecraft at identical times. Uninterrupted communication between the satellites is clearly possible.}
        \label{fig:orbit-scheme}
\end{figure}

The above assumptions reduced the number of dimensions of the parameter space that we need to analyze to just two, which correspond to the perigee and period, equal for both orbits. Now, in order to achieve the desired maximum gravitational potential modulation, the perigee must be set to its minimum possible value. Indeed, for Kepler orbits we have:
\begin{equation}
r_\mathrm{p} = (1-e)a,\quad
r_\mathrm{a} = (1+e)a,\quad
T^2 = \frac{4\pi^2}{GM}a^3,
\label{eq:perigee-apogee-period}
\end{equation}
where $r_\mathrm{p}$ and $r_\mathrm{a}$ are, respectively, the perigee and apogee, $T$ is the period, $a$ is the semi-major axis, and $GM$ is the gravitational parameter of the Earth. It follows from these equations that for each fixed period, $T$, the amplitude of the gravitational potential modulation is maximum when the perigee, $r_\mathrm{p}$, is minimum. We set the perigee to 7500~km, so that the satellites do not enter the Earth's atmosphere. The chosen orbit configuration is depicted in Fig.~\ref{fig:orbit-scheme}. It clearly provides for uninterrupted communication between the satellites, such that the signal propagation paths do not traverse the atmosphere.

With five of the six orbital elements thus fixed (Table~\ref{table:orbital-parameters}), we need to consider the dependence of the experiment accuracy only on the period, equal for both orbits. We will consider the range of period values from 2 to 24~hr. Orbits with smaller periods do not exist for the chosen perigee of 7500~km, while for orbits with larger periods the experiment accuracy decreases, as demonstrated below. We set the experiment duration to 3~years, which is realistic for a space-based experiment. 

\section{Accuracy estimation approach}
\label{sec:accuracy-estimation-approach}

Given the measurements model (\ref{eq:redshift-model}), specification of the clock noise (Table~\ref{table:clock-psds}), satellite orbit parameters (Table~\ref{table:orbital-parameters}), and experiment duration, the problem of estimating the nonrandom EEP violation parameter of interest, $\varepsilon$, is linear and gaussian and may be solved using the maximum likelihood estimator \cite{van-trees-2013}. This approach requires the knowledge of the covariance matrices that correspond to the colored clock noise.

Here we provide explicit expressions for these covariance matrices that we obtained following \cite{williams-2003-jgeod} but without the use of normalization factors introduced there for convenience in geodetic applications. Since $f^{-1}$ and $f^{-2}$ noises are non-stationary, the amplitudes of the corresponding covariance matrices depend on the definition of the PSD, $S(f)$. We use the definition that includes a convolution with the Hanning window \cite{kasdin-95-ieee}. In particular, for a random process $x(t)$ we have:
\begin{equation}
  X(f) = \int_0^T \sin^2(\pi t/T) x(t)e^{-2\pi i t} dt,
\end{equation}
\begin{equation}
  S(f) = \lim_{T\rightarrow \infty}  {1 \over T} {8 \over 3}|X(f)|^2.
\end{equation}
With $S(f)$ thus defined, for the covariance matrices of the three $f^\alpha$ noises of interest, $C^\alpha_{ij}$, we have:
\begin{equation}
C^0_{ij} = \Delta t^{-1} \delta_{ij},
\end{equation}
\begin{equation}
C^{-1}_{ij} = 2\pi \sum^i_{m=1} \phi_m\phi_{i-j+m},\;\;\;\phi_m = {\Gamma (m+1/2) \over m!\, \Gamma (1/2)},
\end{equation}
\begin{equation}
C^{-2}_{ij} = (2 \pi)^2 \Delta t \min(i,j),
\end{equation}
where $\Delta t$ is the sampling interval and $\delta_{ij}$ is the identity matrix.

The experiment model has two unknown non-random parameters, the EEP violation parameter of interest, $\varepsilon$, and the constant frequency shift, $\Delta f_0$, which is a nuisance parameter. We are not interested in the estimation of the value and error of the nuisance parameter of $\Delta f_0$ but, as will be demonstrated below, its presence affects the accuracy of the determination of $\varepsilon$ for experiment durations of order of one orbital revolution. We will characterize the accuracy of $\varepsilon$ by its standard deviation, $\sigma_\varepsilon = \sqrt{\langle\delta\varepsilon^2\rangle}$, and estimate it using the Cramer-Rao bound \cite{van-trees-2013}. Our goal is thus to determine the satellite orbit parameters that minimize $\sigma_\varepsilon$.

\section{Results}
\label{sec:results}

\begin{figure}
    \centering
    \includegraphics[width=0.7\textwidth]{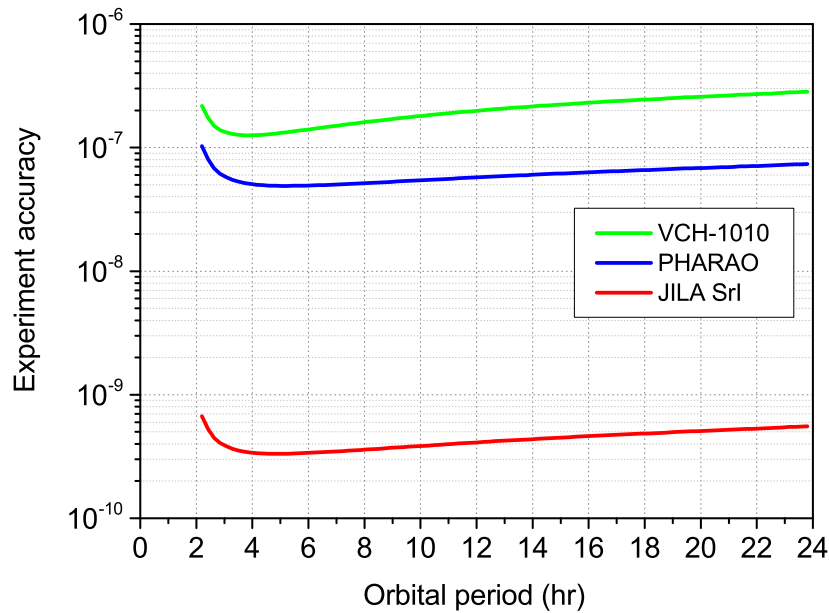}
    \caption{The accuracy of an experiment to test the Einstein Equivalence Principle with two Earth-orbiting satellites as a function of the orbital period. The orbital periods of the two satellites are identical, the values of the other five orbital elements are fixed according to Table~\ref{table:orbital-parameters}. The three considered types of the on-board clocks are: the VCH-1010 hydrogen maser, the PHARAO cesium clock (both space-qualified), and the JILA SrI (laboratory clock). The experiment duration is 3 years.}
        \label{fig:period-accuracy}
\end{figure}

The dependence of the experiment accuracy on the orbital period, estimated using the approach outlined in the previous Section, is presented in Fig.~\ref{fig:period-accuracy}. For the two space-qualified clocks we consider, VCH-1010 and PHARAO, the experiment in its optimal configuration can reach an accuracy of, respectively, $1.3\times10^{-7}$ and $4.9\times10^{-8}$. With the use of an anticipated future space-qualified clock with the performance of the JILA SrI laboratory clock, an accuracy of $3.3\times10^{-10}$ is reachable. This should be compared to the current best results of space-to-ground tests of Gravity Probe A, $1.4\times10^{-4}$ \cite{vessot-levine-1980-prl}, and GREAT, $2.5\times10^{-5}-3.1\times10^{-5}$ \cite{delva-2018-prl, herrmann-2018-prl}.

\begin{figure}
    \centering
    \includegraphics[width=0.7\textwidth]{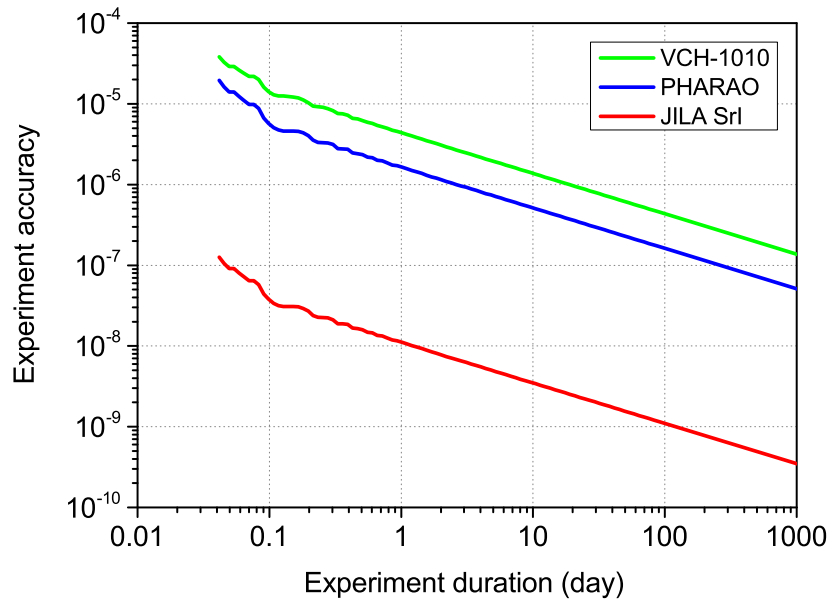}
    \caption{The accuracy of an experiment to test the Einstein Equivalence Principle with two Earth-orbiting satellites as a function of the data accumulation time. The orbital periods of the two satellites are 5~hr, the values of the other five orbital elements are fixed according to Table~\ref{table:orbital-parameters}.}
        \label{fig:time-accuracy}
\end{figure}

An optimal value of the orbital period exists for each of the considered clocks. Deviations from this value of order of a few hours do not affect the accuracy significantly. However, for orbits with periods of order of a day the accuracy decreases significantly: by a factor of 2.3 for VCH-1010, 1.5 for PHARAO, and 1.7 for JILA SrI. The largest accuracy degradation for the VCH-1010 hydrogen maser clock is due to the presence of the Brownian noise which increases rapidly at large averaging times.

The existence of an optimal orbital period, $T\sim$~3--5~hr for modern atomic clocks, appears to be a characteristic feature of the considered experiment type. Indeed, on the one hand, in order to increase the experiment accuracy it is desirable to increase the modulation of the gravitational potential. However, with the perigee fixed, increasing the gravitational potential modulation implies larger orbital periods and thus smaller number of ``repetitions'' of the experiment in a given amount of time. For apogees larger than $\sim50\,000$~km, the gravitational potential difference between the perigee and apogee becomes almost constant, while the clock noise either increases as $T^{1/2}$ (VCH-1010 and other hydrogen masers) or decreases as $T^{-1/2}$ until it hits the ``flicker floor'' (PHARAO, JILA SrI and other cesium and optical clocks). This results in the net decrease of the experiment accuracy for increasing orbital periods. On the other hand, for short orbital periods the number of experiment ``repetitions'' per a given amount of time increases. But this results in the decrease of the gravitational potential modulation, as $T^{2/3}$ for decreasing $T$ (see Eqs.~(\ref{eq:perigee-apogee-period})). The clock noise also increases for short averaging times due to its white component, as $T^{-1/2}$. The net result is, again, the observed decrease of the experiment accuracy for short orbital periods. Therefore, a minimum of $\sigma_\varepsilon$ at some intermediate orbital period necessarily exists.

The dependence of the experiment accuracy on its duration is presented in Fig.~\ref{fig:time-accuracy} for orbits with a fixed period of 5~hr. For large experiment durations, $t$, the uncertainty in $\varepsilon$ falls off as $t^{-1/2}$, as expected. For short experiment durations, significant deviations from this law can be observed. Those are due to the non-sinusoidal form of the gravitational redshift signal and the fact that for accumulation times of order of an orbital period the uncertainties in $\varepsilon$ and $\Delta f_0$ are correlated.

\section{Comparison with a space-to-ground experiment}
\label{sec:space-ground-experiment}

It is of interest to compare the results obtained for the accuracy of a gravitational redshift experiment with two satellites to those of a traditional space-to-ground experiment. Here we do not attempt to approach the problem of determining the optimal configuration of space-to-ground experiments (or find out if it exists). Instead, we pick the RadioAstron gravitational redshift experiment \cite{litvinov-2018-pla} as an example and estimate its accuracy using the model of equation (\ref{eq:redshift-model}). The expected accuracy of this experiment is $\sim2\times10^{-5}$, with the data processing on-going at the time of writing \cite{litvinov-2018-pla, nunes-2020-asr}.

RadioAstron's orbit is not Keplerian but highly evolving due to the gravitational influence of the Moon and other factors. For our analysis we pick the time segment of 01/01/2014--01/01/2015, which starts in the low perigee epoch (perigee altitude $\sim 1000$~km) and ends in the high perigee epoch (respectively, $\sim 50\,000$~km). We use the long-term predicted orbit provided for the RadioAstron mission by the ballistic center of the Keldysh Institute for Applied Mathematics (KIAM) \cite{zakhvatkin-2020-asr}. For the ground station we choose the mission's Green Bank Earth Station with the geodetic latitude of +38$^{\circ}$26$'$16.166$''$, longitude of -79$^{\circ}$50$'$08.810$''$, and height of 812.50~m \cite{langston-2012-nrao-memo}. We assume that both the spaceborne and ground clocks are RadioAstron's VCH-1010 hydrogen maser, with their noise parameters according to Table~\ref{table:clock-psds}. (The hydrogen maser of the ground station actually had a slightly better performance.) For simplicity we assume that the satellite and the ground station are able to communicate continuously. The actual satellite visibility during the considered time interval was approximately half the time.

\begin{figure}
    \centering
    \includegraphics[width=0.7\textwidth]{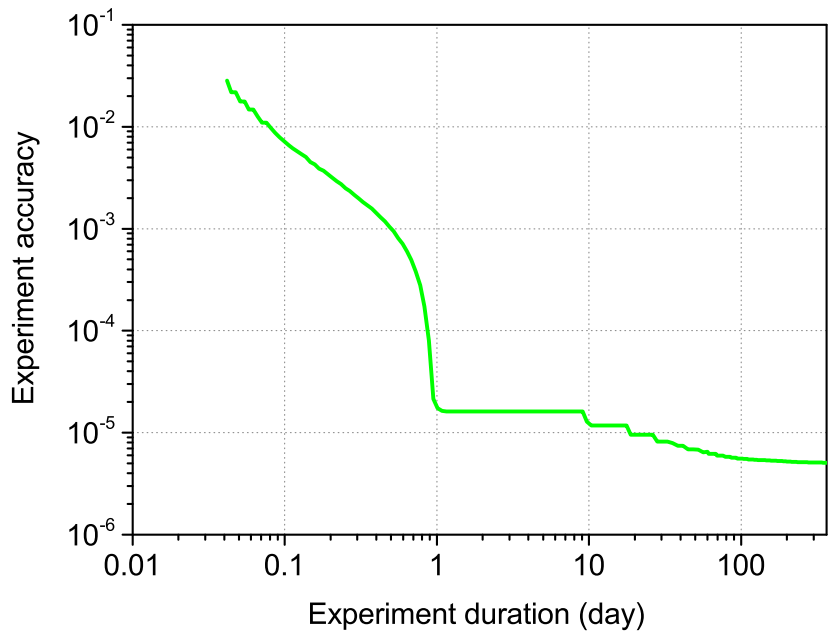}
    \caption{The accuracy of the space-to-ground experiment with the RadioAstron satellite and the Green Bank Earth Station as a function of accumulation time. The simulation starts in the low perigee epoch of 01/01/2014 and ends in the high perigee epoch of 01/01/2015. The accuracy is determined using the simplified assumptions explained in the text.}
        \label{fig:time-accuracy-ra-gbts}
\end{figure}

The accuracy of the experiment with RadioAstron and the Green Bank Earth Station obtained using the model of equation (\ref{eq:redshift-model}) and the above assumptions is presented in Fig.~\ref{fig:time-accuracy-ra-gbts}. After a year of data accumulation the experiment accuracy reaches $5.1\times10^{-6}$, which is more than an order of magnitude lower than that of a space-to-space experiment with the same clocks and the same duration (Fig.~\ref{fig:time-accuracy}). Note that for large accumulation times the accuracy significantly deviates from the $t^{-1/2}$ law. This is due to the orbital evolution mentioned above. Also note the low experiment accuracy for accumulation times of $\sim$~1~day. This is explained by the high correlation between the errors in $\varepsilon$ and $\Delta f_0$ for such accumulation times, which is in turn due to the impulse-like form of the gravitational signal.

If the observational constraints of the real experiment with RadioAstron are taken into account, the above accuracy estimates are decreased by half an order of magnitude \cite{litvinov-2018-pla}.

\section{Conclusions}
\label{sec:conclusions}

We considered the problem of estimating the accuracy of a gravitational redshift test of the Einstein Equivalence Principle that uses two Earth-orbiting satellites equipped with stable atomic clocks. A subclass of the possible orbital configurations, with orbits with identical perigee, apogee and period, has been analyzed. We found out that for such orbits there exists an optimal value of the orbital period that depends on the clock parameters. For the three clocks we considered (VCH-1010, PHARAO, JILA SrI) this value lies in the range of 3--5~hr. The optimal perigee altitude is $\sim 1,000$~km above the Earth's surface.

Using currently available best space-qualified clocks, VCH-1010 and PHARAO, the suggested experiment can reach an accuracy of, respectively, $1\times10^{-7}$ and $5\times10^{-8}$. This is more than 2 orders of magnitude better than the result of the GREAT experiment. An experiment using a space-qualified clock with the performance of the current generation of laboratory optical clocks can reach an accuracy of $3\times10^{-10}$ (for the JILA SrI clock).

The main advantages of the suggested experiment type are the absence of the tropospheric, ionospheric, and antenna mechanical noise in the signal and uninterrupted data accumulation. Here we considered only the random measurement errors inherent to such experiments due to the clock noise. These noises impose the fundamental limit to the accuracy. Another significant error source is the uncertainty in the satellites' orbits. Rough estimates show that, if the contribution of the non-relativistic Doppler effect to the frequency shift is cancelled by a Doppler compensation scheme \cite{vessot-levine-1979-grg}, the orbit determination accuracy required by such experiments can be achieved using the currently available satellite tracking techniques.

\section*{Acknowledgements}

The authors wish to thank Y.~Y.~Kovalev for constant support in the preparation of this paper. We also wish to thank L.~Gurvits for reading the manuscript and making valuable remarks. 
The RadioAstron project is led by the Astro Space Center of the Lebedev Physical Institute of the Russian Academy of Sciences and the Lavochkin Scientific and Production Association under a contract with the Russian Federal Space Agency, in collaboration with partner organizations in Russia and other countries.  

\section*{References}

\bibliographystyle{unsrt}
\bibliography{two_sat_redshift}

\begin{thebibliography}{10}

\bibitem{mcgrew-2018-nat}
W.~F. {McGrew}, X.~{Zhang}, R.~J. {Fasano}, S.~A. {Sch{\"a}ffer}, K.~{Beloy},
  D.~{Nicolodi}, R.~C. {Brown}, N.~{Hinkley}, G.~{Milani}, M.~{Schioppo}, T.~H.
  {Yoon}, and A.~D. {Ludlow}.
\newblock {Atomic clock performance enabling geodesy below the centimetre
  level}.
\newblock {\em Nature}, 564(7734):87--90, November 2018.

\bibitem{bothwell-2019-mlg}
Tobias Bothwell, Dhruv Kedar, Eric Oelker, John~M Robinson, Sarah~L Bromley,
  Weston~L Tew, Jun Ye, and Colin~J Kennedy.
\newblock {JILA} {SrI} optical lattice clock with uncertainty of 2.0e-18.
\newblock {\em Metrologia}, 56(6):065004, oct 2019.

\bibitem{will-2014-lrr}
Clifford~M. Will.
\newblock The confrontation between general relativity and experiment.
\newblock {\em Living Reviews in Relativity}, 17(1):4, 2014.

\bibitem{takamoto-2020-natpho}
Masao {Takamoto}, Ichiro {Ushijima}, Noriaki {Ohmae}, Toshihiro {Yahagi},
  Kensuke {Kokado}, Hisaaki {Shinkai}, and Hidetoshi {Katori}.
\newblock {Test of general relativity by a pair of transportable optical
  lattice clocks}.
\newblock {\em Nature Photonics}, 14(7):411--415, April 2020.

\bibitem{vessot-levine-1980-prl}
R.~F.~C. {Vessot}, M.~W. {Levine}, E.~M. {Mattison}, E.~L. {Blomberg}, T.~E.
  {Hoffman}, G.~U. {Nystrom}, B.~F. {Farrel}, R.~{Decher}, P.~B. {Eby}, and
  C.~R. {Baugher}.
\newblock {Test of relativistic gravitation with a space-borne hydrogen maser}.
\newblock {\em Physical Review Letters}, 45:2081--2084, December 1980.

\bibitem{delva-2018-prl}
P.~{Delva}, N.~{Puchades}, E.~{Sch{\"o}nemann}, F.~{Dilssner}, C.~{Courde},
  S.~{Bertone}, F.~{Gonzalez}, A.~{Hees}, Ch. {Le Poncin-Lafitte},
  F.~{Meynadier}, R.~{Prieto-Cerdeira}, B.~{Sohet}, J.~{Ventura-Traveset}, and
  P.~{Wolf}.
\newblock {Gravitational Redshift Test Using Eccentric Galileo Satellites}.
\newblock {\em Phys. Rev. Lett.}, 121(23):231101, Dec 2018.

\bibitem{herrmann-2018-prl}
Sven {Herrmann}, Felix {Finke}, Martin {L{\"u}lf}, Olga {Kichakova}, Dirk
  {Puetzfeld}, Daniela {Knickmann}, Meike {List}, Benny {Rievers}, Gabriele
  {Giorgi}, Christoph {G{\"u}nther}, Hansj{\"o}rg {Dittus}, Roberto
  {Prieto-Cerdeira}, Florian {Dilssner}, Francisco {Gonzalez}, Erik
  {Sch{\"o}nemann}, Javier {Ventura-Traveset}, and Claus {L{\"a}mmerzahl}.
\newblock {Test of the Gravitational Redshift with Galileo Satellites in an
  Eccentric Orbit}.
\newblock {\em Phys. Rev. Lett.}, 121(23):231102, December 2018.

\bibitem{litvinov-2018-pla}
D.~A. {Litvinov}, V.~N. {Rudenko}, A.~V. {Alakoz}, U.~{Bach}, N.~{Bartel},
  A.~V. {Belonenko}, K.~G. {Belousov}, M.~{Bietenholz}, A.~V. {Biriukov},
  R.~{Carman}, G.~{Cim{\'o}}, C.~{Courde}, D.~{Dirkx}, D.~A. {Duev}, A.~I.
  {Filetkin}, G.~{Granato}, L.~I. {Gurvits}, A.~V. {Gusev}, R.~{Haas},
  G.~{Herold}, A.~{Kahlon}, B.~Z. {Kanevsky}, V.~L. {Kauts}, G.~D.
  {Kopelyansky}, A.~V. {Kovalenko}, G.~{Kronschnabl}, V.~V. {Kulagin}, A.~M.
  {Kutkin}, M.~{Lindqvist}, J.~E.~J. {Lovell}, H.~{Mariey}, J.~{McCallum},
  G.~{Molera Calv{\'e}s}, C.~{Moore}, K.~{Moore}, A.~{Neidhardt},
  C.~{Pl{\"o}tz}, S.~V. {Pogrebenko}, A.~{Pollard}, N.~K. {Porayko},
  J.~{Quick}, A.~I. {Smirnov}, K.~V. {Sokolovsky}, V.~A. {Stepanyants}, J.~M.
  {Torre}, P.~{de Vicente}, J.~{Yang}, and M.~V. {Zakhvatkin}.
\newblock {Probing the gravitational redshift with an Earth-orbiting
  satellite}.
\newblock {\em Physics Letters A}, 382(33):2192--2198, Aug 2018.

\bibitem{origlia-2018-pra}
S.~{Origlia}, M.~S. {Pramod}, S.~{Schiller}, Y.~{Singh}, K.~{Bongs},
  R.~{Schwarz}, A.~{Al-Masoudi}, S.~{D{\"o}rscher}, S.~{Herbers},
  S.~{H{\"a}fner}, U.~{Sterr}, and Ch. {Lisdat}.
\newblock {Towards an optical clock for space: Compact, high-performance
  optical lattice clock based on bosonic atoms}.
\newblock {\em Physical Review A}, 98(5):053443, November 2018.

\bibitem{vessot-levine-1979-grg}
R.~F.~C. {Vessot} and M.~W. {Levine}.
\newblock {A test of the equivalence principle using a space-borne clock}.
\newblock {\em General Relativity and Gravitation}, 10:181--204, February 1979.

\bibitem{aces-2011-acau}
M.~P. {He{\ss}}, L.~{Stringhetti}, B.~{Hummelsberger}, K.~{Hausner},
  R.~{Stalford}, R.~{Nasca}, L.~{Cacciapuoti}, R.~{Much}, S.~{Feltham},
  T.~{Vudali}, B.~{L{\'e}ger}, F.~{Picard}, D.~{Massonnet}, P.~{Rochat},
  D.~{Goujon}, W.~{Sch{\"a}fer}, P.~{Laurent}, P.~{Lemonde}, A.~{Clairon},
  P.~{Wolf}, C.~{Salomon}, I.~{Proch{\'a}zka}, U.~{Schreiber}, and
  O.~{Montenbruck}.
\newblock {The ACES mission: System development and test status}.
\newblock {\em Acta Astronautica}, 69:929--938, December 2011.

\bibitem{smarr-vessot-1983-grg}
L.~L. Smarr, R.~F.~C. Vessot, C.~A. Lundquist, R.~Decher, and Tsvi Piran.
\newblock Gravitational waves and red shifts: A space experiment for testing
  relativistic gravity using multiple time-correlated radio signals.
\newblock {\em General Relativity and Gravitation}, 15(2):129--163, 1983.

\bibitem{tinto-1996-prd}
Massimo {Tinto}.
\newblock {Spacecraft Doppler tracking as a xylophone detector of gravitational
  radiation}.
\newblock {\em Phys. Rev. D}, 53(10):5354--5364, May 1996.

\bibitem{altschul-2015-asr}
Brett {Altschul}, Quentin~G. {Bailey}, Luc {Blanchet}, Kai {Bongs}, Philippe
  {Bouyer}, Luigi {Cacciapuoti}, Salvatore {Capozziello}, Naceur {Gaaloul},
  Domenico {Giulini}, Jonas {Hartwig}, Luciano {Iess}, Philippe {Jetzer},
  Arnaud {Land ragin}, Ernst {Rasel}, Serge {Reynaud}, Stephan {Schiller},
  Christian {Schubert}, Fiodor {Sorrentino}, Uwe {Sterr}, Jay~D. {Tasson},
  Guglielmo~M. {Tino}, Philip {Tuckey}, and Peter {Wolf}.
\newblock {Quantum tests of the Einstein Equivalence Principle with the
  STE-QUEST space mission}.
\newblock {\em Advances in Space Research}, 55(1):501--524, Jan 2015.

\bibitem{wolf-blanchet-2016-cqg}
Peter {Wolf} and Luc {Blanchet}.
\newblock {Analysis of Sun/Moon gravitational redshift tests with the STE-QUEST
  space mission}.
\newblock {\em Classical and Quantum Gravity}, 33(3):035012, Feb 2016.

\bibitem{van-trees-2013}
H.~L. {van Trees}, K.~L. {Bell}, and Z.~{Tian}.
\newblock {\em Detection, Estimation, and Modulation Theory. Part 1 -
  Detection, Estimation, and Filtering Theory}.
\newblock Wiley, New York, USA, 2nd edition, 2013.

\bibitem{williams-2003-jgeod}
Simon Williams.
\newblock The effect of coloured noise on the uncertainties of rates estimated
  from geodetic time series.
\newblock {\em Journal of Geodesy}, 76:483--494, 02 2003.

\bibitem{kasdin-95-ieee}
N.~J. {Kasdin}.
\newblock Discrete simulation of colored noise and stochastic processes and
  1/f/sup /spl alpha// power law noise generation.
\newblock {\em Proceedings of the IEEE}, 83(5):802--827, 1995.

\bibitem{montenbruck-2013-gpss}
Oliver Montenbruck, Andr\'{e} Hauschild, Yago Andres, Axel Engeln, and
  Christian Marquardt.
\newblock (near-)real-time orbit determination for gnss radio occultation
  processing.
\newblock {\em GPS Solut.}, 17(2):199–209, April 2013.

\bibitem{montenbruck-2018-gpss}
Oliver Montenbruck, Stefan Hackel, Jose Ijssel, and Daniel Arnold.
\newblock Reduced dynamic and kinematic precise orbit determination for the
  swarm mission from 4 years of gps tracking.
\newblock {\em GPS Solut.}, 22(3):1–11, July 2018.

\bibitem{nunes-2020-asr}
N.~V. {Nunes}, N.~{Bartel}, M.~F. {Bietenholz}, M.~V. {Zakhvatkin}, D.~A.
  {Litvinov}, V.~N. {Rudenko}, L.~I. {Gurvits}, G.~{Granato}, and D.~{Dirkx}.
\newblock {The gravitational redshift monitored with RadioAstron from near
  Earth up to 350,000 km}.
\newblock {\em Advances in Space Research}, 65(2):790--797, January 2020.

\bibitem{zakhvatkin-2020-asr}
M.~V. {Zakhvatkin}, A.~S. {Andrianov}, V.~Yu. {Avdeev}, V.~I. {Kostenko}, Y.~Y.
  {Kovalev}, S.~F. {Likhachev}, I.~D. {Litovchenko}, D.~A. {Litvinov}, A.~G.
  {Rudnitskiy}, M.~A. {Shchurov}, K.~V. {Sokolovsky}, V.~A. {Stepanyants},
  A.~G. {Tuchin}, P.~A. {Voitsik}, G.~S. {Zaslavskiy}, V.~E. {Zharov}, and
  V.~A. {Zuga}.
\newblock {RadioAstron orbit determination and evaluation of its results using
  correlation of space-VLBI observations}.
\newblock {\em Advances in Space Research}, 65(2):798--812, January 2020.

\bibitem{langston-2012-nrao-memo}
G.~{Langston}.
\newblock Nrao 43m antenna coordinates and angular limits.
\newblock Technical Report EDIR Memo \#324, National Radio Astronomy
  Observatory, Charlottesville, Virginia, 2012.

\end{thebibliography}

\end{document}